\documentclass[letterpaper,10pt]{article}
\usepackage{setspace}
\usepackage{natbib}
\usepackage{amsfonts}
\usepackage{bm}
\usepackage{graphicx}
\usepackage{amsmath,amsfonts,amssymb,amsthm}
\usepackage{xcolor}
\usepackage{float}
\usepackage[T1]{fontenc}
\usepackage[utf8]{inputenc}
\usepackage[english]{babel}
\usepackage{authblk}
\restylefloat{table}
\usepackage[left=2.54cm,top=2.54cm,right=2.54cm,bottom=2.54cm]{geometry}
\title{A Hierarchical Spatio-Temporal Statistical Model Motivated by Glaciology}
\author[1]{Giri Gopalan}
\author[1]{Birgir Hrafnkelsson}
\author[2]{Christopher K. Wikle}
\author[3]{H\aa vard Rue}
\author[1]{Guðfinna Aðalgeirsd\'ottir}
\author[4]{Alexander H. Jarosch}
\author[1]{Finnur P\'alsson}
\begin{document}
\affil[1]{University of Iceland. Reykjavik, Iceland. 101.}
\affil[2]{University of Missouri. Columbia, Missouri. 65211.}
\affil[3]{King Abdullah University of Science and Technology. Thuwal, Saudi Arabia. 23955.}
\affil[4]{University of Innsbruck. Innsbruck, Austria. 6020.}

\renewcommand\Authands{ and }

\date{}

\maketitle

\date{}

\maketitle

\begin{abstract}
In this paper, we extend and analyze a Bayesian hierarchical spatio-temporal model for physical systems. A novelty is to model the discrepancy between the output of a computer simulator for a physical process and the actual process values with a multivariate random walk. For computational efficiency, linear algebra for bandwidth limited matrices is utilized, and first-order emulator inference allows for the fast emulation of a numerical partial differential equation (PDE) solver. A test scenario from a physical system motivated by glaciology is used to examine the speed and accuracy of the computational methods used, in addition to the viability of modeling assumptions. We conclude by discussing how the model and associated methodology can be applied in other physical contexts besides glaciology.
\end{abstract}


\newpage
\section{Introduction}

Scientists and engineers often study a physical system with the goal of making spatio-temporal predictions (e.g., temperature or glacier thickness) and inferring unknown quantities governing the system (e.g., atmospheric density or ice viscosity). This system's dynamics can often be phrased in terms of spatio-temporal partial differential equations (PDEs) that are based on approximations. The scientist or engineer may also be able to simulate the physical system with a computer simulator, such as a numerical PDE solver, which is subject to imperfections (e.g., numerical error). Moreover, the scientific constants entering into the system's dynamical equations such as density, friction, or viscosity may not be known precisely, but their range can be constrained to some set of plausible values. Additionally field data, though potentially scarce and noisy, can be incorporated into the analysis.

Such scenarios can be modeled with a variant of a Bayesian hierarchical spatio-temporal model that was introduced in \citet{gopalan2018bayesian} for glacial dynamics, if considered more generally. We delineate three methods to make posterior inference efficient: the first is to utilize bandwidth limited linear-algebraic routines for likelihood evaluation \citep{rue2001fast}, the second is to utilize an embarrassingly parallel approximation to the likelihood, and the third is to use first-order emulators \citep{Hooten2011} for speeding up computer simulators. Though our modeling and numerical results are still within a glaciology context, we conclude with a discussion of how the model can be applied to other physical scenarios. Before introducing the Bayesian hierarchical model and associated methodology for computationally efficient posterior inference, it is appropriate to summarize relevant statistical literature developed over the last two decades.

Bayesian hierarchical modeling for geophysical problems was introduced in \citet{10.1007/978-94-011-5430-7_3} and \citet{Wikle1998}, and summarized in \citet{Berliner}, \citet{cressie2011statistics}, and \citet{Wikle2016}. In this modeling approach, prior distributions are specified for physical parameters of interest, a physical process is modeled at the intermediary, latent level (conditional on the physical parameters), and the data collection process is modeled conditional on the latent physical process values. Both numerical error and model uncertainty can be incorporated at the process level, while measurement errors can be modeled at the data level. This approach has been applied in a variety of scientific contexts, including the study of ozone concentrations \citep{berrocal2014assessing}, sediment loads at the Great Barrier Reef \citep{pagendam2014assimilating}, precipitation in Iceland \citep{RePEc:wly:envmet:v:27:y:2016:i:1:p:27-41}, Antarctic contributions to sea level rise \citep{zammit2014resolving}, and tropical ocean surface winds \citep{wikle2001spatiotemporal} (among many others). In \citet{gopalan2018bayesian}, the motivating example for the work in this paper, a Bayesian hierarchical model for shallow glaciers based on the shallow ice approximation (SIA) PDE was developed and evaluated. 

\citet{kennedy2001bayesian} suggest constructing Bayesian statistical models that incorporate the output of a computer simulator of a physical process, such as a numerical solver for the underlying system of PDEs. Fundamental to their approach is the inclusion of a specific term that represents the deviation between the output of a computer simulator and the actual process values, known as \textit{model discrepancy} or \textit{model inadequacy}. This framework is developed in \citet{higdon2004combining}, \citet{higdon2008computer}, and \citet{discrepancy}. In particular, \citet{higdon2008computer} use a Bayesian model along with a principal components based approach for reducing the computational overhead of running a computer simulation with high dimensional output multiple times (an approach termed as \textit{emulation}). \citet{discrepancy} note that the prior for model discrepancy must be chosen carefully to mitigate bias of physical parameters and predictions. In particular, as more prior information is incorporated into a model discrepancy term through a constrained Gaussian process (GP) prior over a space of functions, the less biased inferences and predictions tend to become. The notions of an emulator, a computer simulator, and model discrepancy enter naturally into the aforementioned Bayesian hierarchical framework. Conditional on physical parameters coupled with initial and/or boundary conditions, the physical process values at the latent level can be written as the sum of a computer simulator or emulator term and a model discrepancy term. 

To be precise, let us assume that the physical process $\bm{S}$ can be indexed through time, i.e., as $\bm{S}_j$, and $\bm{S}_j$ is a vector where each element corresponds to a distinct spatial location. One can specify the process level conditional on physical parameter $\bm{\theta}$ as 
\begin{eqnarray}
\bm{S}_j &=& \bm{f}(\bm{\theta},j)+\bm{\delta}(j)
\end{eqnarray}
where $\bm{\delta}(.)$ is a vector valued model discrepancy function that is independent of $\bm{\theta}$, and $\bm{f}(\bm{\theta},j)$ is the output of a computer simulation or emulator for physical parameter $\bm{\theta}$ at time index $j$. If, for instance, at each time point $j$ an observation $\bm{Y}_j$ of $\bm{S}_j$ is made with associated measurement error $\bm{\eta}_j$, then observations can be written as
\begin{eqnarray}
\bm{Y}_j &=& \bm{f}(\bm{\theta},j)+\bm{\delta}(j)+\bm{\eta}_j,
\end{eqnarray}
which is analogous to Eq. 5 of \citet{kennedy2001bayesian}.

In \citet{kennedy2001bayesian}, $\bm{\delta}(.)$ is a fixed but unknown function independent of $\bm{\theta}$ that is learned with a GP prior distribution. Similarly, $\bm{\delta}(.)$ has a constrained GP prior in \citet{discrepancy}. The approach in this paper instead assumes a temporally indexed stochastic process (with spatial correlation) that follows a multivariate random walk, rather than a deterministic function. Additionally, in \citet{liu2009}, the authors frame a computer emulator of time series run under multiple inputs as a dynamic linear model (DLM). As part of their approach, they allow for time varying auto-regressive coefficients that follow a random walk process, to embed non-stationarity into the model.

While the approach taken in this paper most closely follows the above literature (i.e., Bayesian hierarchical modeling, model discrepancy, and emulation), we briefly review literature in probabilistic numerics and Bayesian numerical analysis; the emphasis in Bayesian numerical analysis is to use probabilistic methods to solve numerical problems, whereas, in the Bayesian hierarchical setup, one is also interested in inference of scientifically relevant parameters and predictions of the physical process. In \citet{conrad2017statistical}, a probabilistic ordinary differential equation (ODE) solver is developed that adds stochasticity at each iteration; conditions for the convergence of this method to the ODE solution are given. \citet{chkrebtii2016bayesian} utilize GPs for solving ODEs; moreover, \citet{Calderhead:2008:ABI:2981780.2981808} use a GP regression based method to avoid explicitly solving nonlinear ODEs when performing inference for parameters that provides computational speed ups; additionally, \citet{Owhadi} present a gamblet based solver that comes with provably computationally efficient solutions to PDEs. The approach is derived from a game theoretic and stochastic PDE framework.

In the spatio-temporal model described in this paper, stochasticity is induced with an error-correcting process that is separated from the numerical solution. In general, another way to achieve this is to define a stochastic process by equating a PDE to a white noise term -- that is, the solution $\bm{X}$ to a stochastic partial differential equation (SPDE) $L[\bm{X}] = \bm{W}$, where $\bm{L}$ is a differential operator and $\bm{W}$ is a white noise process (indexed by spatio-temporal coordinates). For instance, a fractional Laplacian operator yields the Mat\'{e}rn covariance function \citep{whittle1954stationary,Whittle63,lindgren2011explicit}. We employ the former approach mainly because it is difficult to derive exact covariance functions for arbitrary differential equations (e.g., in the presence of nonlinearities), though we highlight the utility of the latter approach in situations where an analytical covariance function can be derived exactly.

 A major feature of this work is to represent the discrepancy between real physical process values and the output of a computer simulator for these physical process values as a multivariate random walk; typically, model discrepancy is endowed with a GP prior or a constrained GP prior over a space of functions as in \citet{kennedy2001bayesian} and \citet{discrepancy}. Along with this model is the development of two ways for making computations faster: the first is harnessing first-order emulator inference \citep{Hooten2011} for speeding up the computation of a numerical solver, and the second is the use of bandwidth limited numerical linear algebra \citep{rue2001fast} for computing the likelihood efficiently. Moreover, in the regime of a high signal-to-noise ratio, an embarrassingly parallel approximation to the likelihood can be employed. Finally, methodology to fit a spatial Gaussian field for the log of the scale of numerical errors is discussed.
 
We must also be clear about what distinguishes this work from its predecessor, \citet{gopalan2018bayesian}. This includes the use of emulators, probing higher order random walks besides order 1, derivation of sparsity and computational complexity of log-likelihood evaluation, empirical run time results, and methodology to fit an error-correcting process when little prior information is available. The structure of this paper is as follows: First a test system from glaciology is described. Then the statistical model that is the focus of this work is presented in detail (in the context of the glaciology test case), followed by the exact and approximate likelihood. Then this model is analyzed in terms of computational run time and accuracy of inference, based on the test system from glaciology; moreover, the random walk error-correcting process is assessed with residual analysis. Afterward, we discuss how the model and associated methodology can be applied to other physical scenarios, and conclude by summarizing the model, method, and limitations of the approach.

 \section{Description of a test system from glaciology}
Before delving into the specifics of the Bayesian hierarchical model and computational subtleties, we begin with a brief discussion of glaciology. Glaciology is the study of physical systems consisting mostly of ice and snow. This broad definition includes the study of the crystalline nature of ice, the transformation and compaction of snow into ice, the dynamics of the flow of ice and water in a glacier, the relationships between fundamental quantities like viscosity, temperature, and pressure, the relationships between precipitation and meteorology with said ice systems, the interaction of ice systems with other geological systems such as volcanoes and bedrock, and so on. As such, glaciology synthesizes elements from a multitude of scientific disciplines including continuum mechanics, fluid mechanics, hydraulics, chemistry, and meteorology.

\cite{Bueler} introduce analytical solutions for the SIA PDE, a commonly used model for the dynamics of glaciers \citep{10.2307/79748, doi:10.1080/03091928208209013, flowers2005sensitivity,Paterson,vanderveen, Brinkerhoff, 2016arXiv161201454G, gopalan2018bayesian}. Based on the principle of conservation of mass, the SIA dictates that glacier flow is in the direction of the (negative) gradient of the glacier surface and is due to gravity and basal sliding (also referred to as friction or drag if in the direction of the positive gradient). While an explanation of the SIA PDE is given in \cite{gopalan2018bayesian}, our focus is on ice viscosity, $B$. Intuitively, this parameter controls the softness of the ice. The other main physical parameter, which is not the subject of this paper, is $C_0\gamma$. This controls basal sliding or friction. 

For the analysis that follows, we focus on a periodic solution to the SIA in which the thickness of the glacier oscillates through time; $H(r,t)$, the thickness of the glacier as a function of two dimensional space (in polar coordinates) and time, is 
\begin{eqnarray}
H(r,t) &=& H_s(r)+P(r,t), \\
P(r,t)  &=& C_p\sin(2\pi t/T_p)\cos^2\left[\frac{\pi(r-0.6L)}{.6L}\right]; \textrm{if } 0.3L < r < .9L, \\
P(r,t) &=& 0; \textrm{if } 0 \leq r \leq 0.3L \textrm{ or if } r \geq 0.9L. 
\end{eqnarray}
In Eq. 3, $H_s$ is a static initial profile of the glacier (i.e., a dome as in Eq. 21 of \citet{Bueler}), $P$ is a perturbation (e.g., precipitation) function, $L$ is the margin length, $C_p$ is the magnitude of the periodic perturbation, and $T_p$ is the period of the perturbation. \citet{Bueler} derive a mass balance function that achieves this periodic solution for the SIA PDE. Qualitatively, this test case appears like a dome with a periodic oscillation in thickness around an annulus defined by $0.3L < r < .9L$. In Figure 1, an illustration of the oscillations of glacier thickness through time is displayed.

The value of each surface elevation measurement is the value of the exact analytical solution above added to a zero-mean Gaussian random variable with standard deviation of 1 meter, larger than errors of the digital-GPS instruments employed by the UI-IES. We use the same values of parameters as in \citet{Bueler} to make for easier comparison to that work and the EISMINT experiment. In particular, $H_0 = 3600$ m, $L = 750$ km, $C_p = 200$ m, and $T_p = 5000$ years.

Employing the same set up as \citet{gopalan2018bayesian}, glacial surface elevation measurements are assumed to be collected for 20 years, twice a year, and at 25 fixed spatial locations across the glacier, to emulate how the glaciology team at the University of Iceland Institute of Earth Sciences (UI-IES) collects data at Icelandic glaciers (e.g., see Figure 2 illustrating Langj\"{o}kull and the mass balance measurement sites). 

\begin{figure*}[h!]
  \centering
    \includegraphics[width=0.5\textwidth]{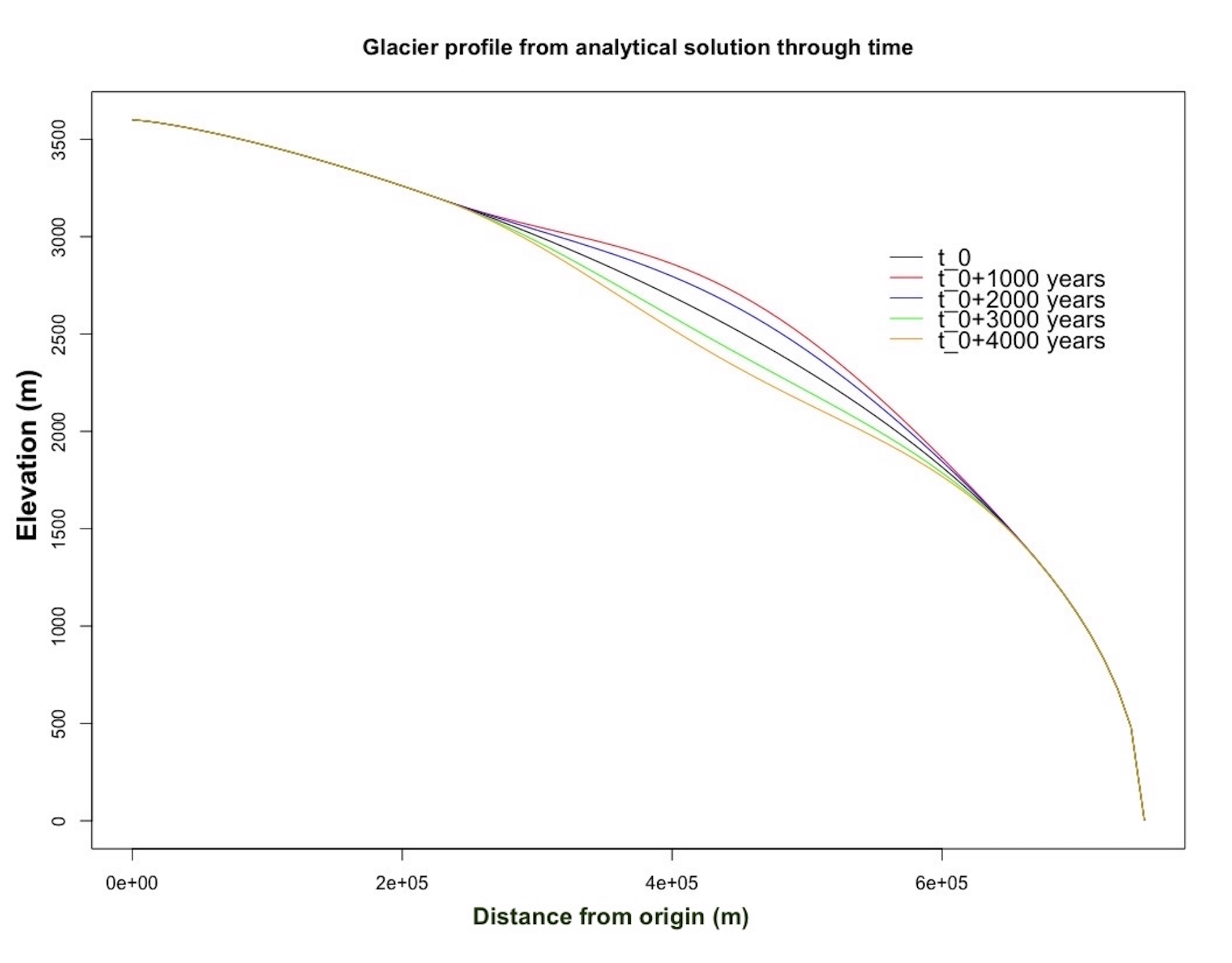}
\caption{An illustration of the periodic oscillatory exact solution to the SIA PDE that is used for the analysis. Since the solution is radially symmetric, only a radial cross section is illustrated. This solution is stationary except for an annulus defined by $0.3L < r < .9L$, where $L$ is 750 $km$; in the annulus, the glacier thickness vibrates back and forth periodically, as illustrated.}
\end{figure*}

\begin{figure*}[h!]
  \centering
    \includegraphics[width=0.5\textwidth]{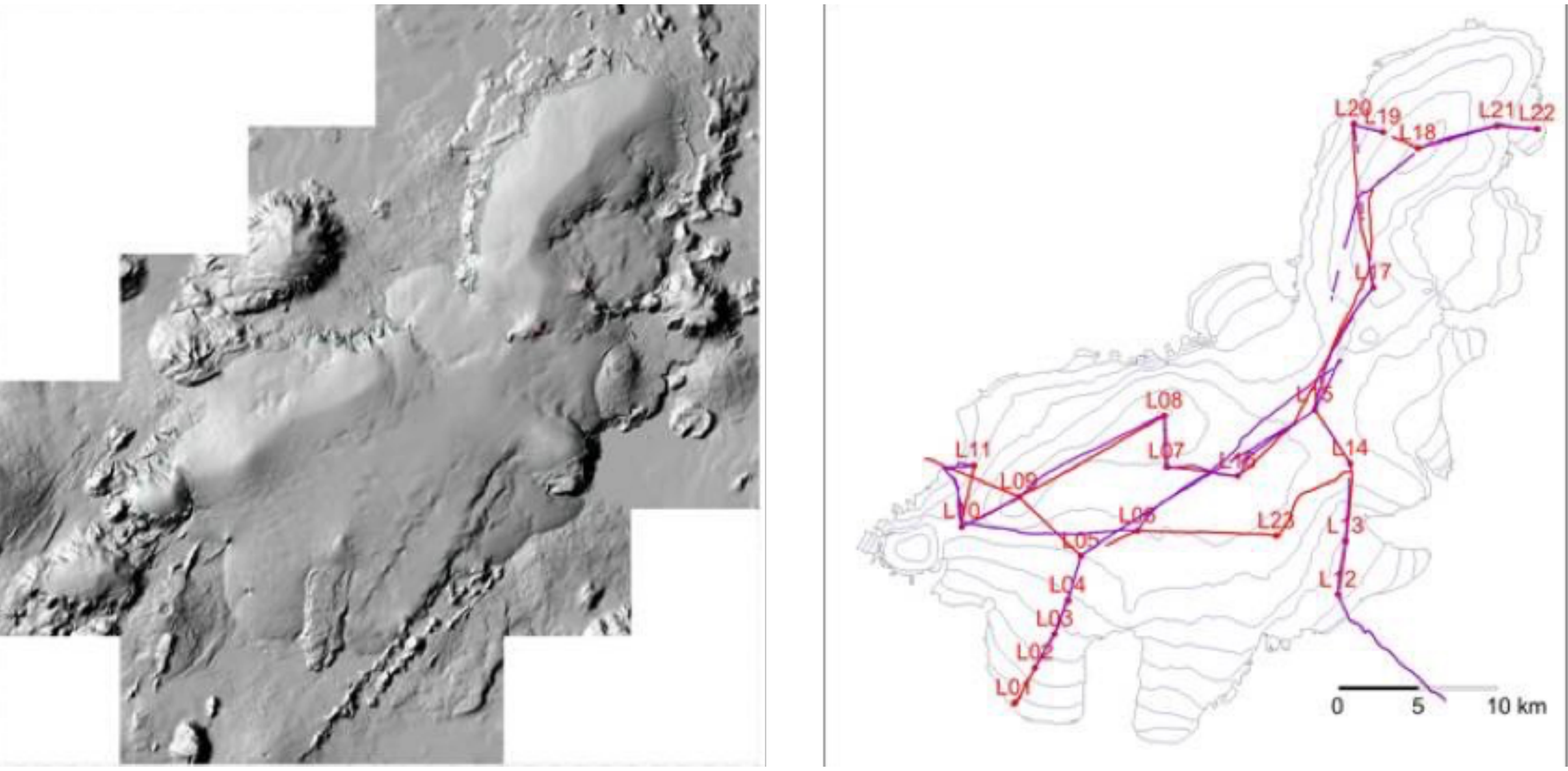}
\caption{A digital elevation map of Langj\"{o}kull along with measurement sites demarcated on the right, provided by the University of Iceland Institute of Earth Sciences (UI-IES). Langj\"{o}kull is Iceland's second largest glacier by area, at $900$ sq. km, and its mean thickness is 210 meters above sea level \citep{bjornsson2008icelandic}, so Langj\"{o}kull is shallow.}
\end{figure*}

\section{The hierarchical spatio-temporal model and its properties}
Now that we have acquainted the reader with some facts about glaciology and the particular test case used for the analysis in this paper, we next delineate the hierarchical spatio-temporal model that is the focus of this work, by specifying its variables, parameters, and properties, including efficient computation of the likelihood and connections to other modeling frameworks. For the sake of specificity of the presentation, the glaciology example is referenced, similarly to the set up in \citet{gopalan2018bayesian}. We assume that $n$ spatial locations are modeled at the latent level, and $m$ of those locations are observed, where $m$ is typically much smaller than $n$. We use the index $j$ to refer to time indices and $i$ to refer to spatial indices; while space and time are discretized, the differences between successive time and spatial points can be made as small as desired depending on the context of the application and computational resources available. Throughout, we use bolded notation for vectors and uppercase, unbolded, and non-italic notation for matrices. All other mathematical symbols are scalars.

We introduce the Bayesian hierarchical model in the parameter, process, data level framework of \citet{10.1007/978-94-011-5430-7_3}. We denote the physical parameters as $\bm{\theta}$ and initial and/or boundary conditions for the physical process as $\bm{\phi}$. At the parameter level, one possibility is to use a truncated normal distribution for $\bm{\theta}$ if the support of the parameter value can be constrained, as was done in \citet{gopalan2018bayesian}, where $\bm{\theta}$ represented ice viscosity. However, more generally, the distribution can be specified based on domain knowledge or expertise. We denote the output of a computer simulator, which could be either a numerical solver or an emulator, at time $j$ with the notation $\bm{f}(\bm{\theta},\bm{\phi}, j)$, which, in full generality, is an element of $\mathbb{R}^n$. While some values could be negative (e.g., temperature), in many cases the computer simulator output can be restricted to the nonnegative real numbers. For a specific example, in Appendix A of \citet{gopalan2018bayesian}, $\bm{f}(\bm{\theta},\bm{\phi}, j)$ is a second-order finite difference solver for glacier thickness, which is constrained to be nonnegative based on a boundary condition. Evidence for a nonnegative support for the physical process, in glaciology, can be found in \cite{gopalan2018bayesian}. Particularly, this is evident in Figure 6 of that paper, which shows the process (i.e., glacier thickness) predictions across the glacier, and the distributions are all greater than zero. Specifically, the minimum of the smallest box-plot is more than 750 m. Nonetheless, the reader is suggested to think carefully about whether a negligible amount of probability mass is below zero in different applications (e.g., temperature models).

The process level of the model, conditional on $\bm{\theta}$ and $\bm{\phi}$, can be written as:

\begin{eqnarray}
\bm{X}_{j} &=& \bm{X}_{j-1} +\bm{\epsilon}_{j}, \\
\bm{S}_{j} &=& \bm{f}(\bm{\theta},\bm{\phi},j) + \bm{X}_j,
\end{eqnarray}
where $\bm{X}_{0}$ is a vector of zeros.

In the above expressions, $\bm{\epsilon}_{j}$ is $\textrm{MVN}(0,\Sigma)$ and independent of $\bm{\epsilon}_{l}$ for $j \neq l$. Furthermore, $\bm{X}_{j}$, $\bm{\epsilon}_{j}$,  $\bm{f}(\bm{\theta},\bm{\phi},j)$, and (consequently) $\bm{S}_{j}$ are members of $\mathbb{R}^n$. In \citet{gopalan2018bayesian},  $\{\bm{X}_1, \bm{X}_2, ..., \}$ was referred to as an error-correcting process because it was meant to represent the difference between the numerical solver and  the exact solution to the SIA PDE. Note that in \citet{gopalan2018bayesian}, $\bm{S}_j$ referred to glacial thickness at a particular time point, where each component referred to the glacial thickness at a particular grid point. In more generality, the error-correcting statistical process can be a random walk of higher order; a multivariate RW process of order $q$ ($RW(q)$) is given by:
\begin{eqnarray}
\bm{X}_{j}+ \sum_{p=1}^q(-1)^{p}{q \choose p}\bm{X}_{j-p} &=& \bm{\epsilon}_j
\end{eqnarray}
where $\bm{\epsilon}_1$,..., $\bm{\epsilon}_q$ are independent and marginally $\textrm{MVN}(0,\Sigma)$. This form of a higher order random walk is a multivariate extension of the integrated auto-regressive process given in Chapter 5.6 of \citet{madsen2007time}. For $q=2$, this corresponds to RW(2) of \citet{rue2005gaussian}.

At the data level, it is assumed that data are regularly sampled at every $k$-th time point, so that one observes $\bm{Y}_k, \bm{Y}_{2k},..., \bm{Y}_{Nk} \in \mathbb{R}^m$; in the glaciology test case, the variables $\bm{Y}$ referred to glacial surface elevation measurements, and $k$ was set to 5, to represent the fact that the glaciologists take a set of measurements in the summer and winter, or twice a year. The corresponding observation errors  $\bm{\eta}_k, \bm{\eta}_{2k},..., \bm{\eta}_{Nk}$ are IID $\textrm{MVN}(0,\sigma^2\textrm{I})$, and represent digitial-GPS measurement errors in the glaciology example. We define the matrix  A $\in \mathbb{R}^{m \times n}$ to be such that its rows are unit basis vectors (i.e., an incidence matrix as in \citet{cressie2011statistics}). That is, $\textrm{A}_{ab} = 1$ if and only if the $b$th index of the process level vector $\bm{S}$ has been observed, and $\textrm{A}_{ab} = 0$ for all other entries. Then the data level model, conditional on the process $\bm{S}$, is

\begin{eqnarray}
\bm{Y}_{ck} &=& \textrm{A}\bm{S}_{ck}+\bm{\eta}_{ck}, 
\end{eqnarray}
where we assume that  $j \in \{1,2,..., T\}$ and $c \in \{1,2,..., N\}$, so there are $N$ total observed spatial vectors, observed with a period of length $k$. 

Conditional on $\bm{\theta}$, $\bm{\phi}$, and a computer simulator, the model can be thought of as a hidden Markov model (HMM) \citep{baum1966}; the latent physical process evolves according to a RW(1) process added to a numerical solution, and it is observed indirectly with Gaussian noise. It can also be thought of as a conditional general state space model. This is because, conditioning on  $\bm{\theta}$, $\bm{\phi}$, and a computer simulator, one can write:

\begin{eqnarray}
\bm{S}_{j} &=& \bm{S}_{j-1}+[-\bm{f}(\bm{\theta},\bm{\phi},j-1)+\bm{f}(\bm{\theta},\bm{\phi},j)] + \bm{\epsilon}_j,\\
\bm{Y}_{ck} &=& \textrm{A}\bm{S}_{ck}+\bm{\eta}_{ck}.
\end{eqnarray}
Here, the state evolves linearly with a time dependent offset term: $[-\bm{f}(\bm{\theta},\bm{\phi},j-1)+\bm{f}(\bm{\theta},\bm{\phi},j)]$. The notation $ck$ is used in Eq. 11 to indicate that observations of the process are only observed every $k$th time point, whereas the latent process evolves at every time step $j$. The reader who is interested in further understanding the connection between Gaussian processes and state space models may consult \cite{pmlr-v33-solin14}. 

\subsection{Exact likelihood}
An advantage of using this model is that the likelihood, $p(\bm{Y}_k, \bm{Y}_{2k},..., \bm{Y}_{Nk}|\bm{\theta},\bm{\phi})$, can be computed exactly in an efficient manner. It can also be approximated in a way that leads to embarrassingly parallel computation when the signal-to-noise ratio is high. The next several sections provide more details for these considerations. The likelihood of the model, $L(\bm{\theta},\bm{\phi}) = p(\bm{Y}_k, \bm{Y}_{2k},..., \bm{Y}_{Nk}|\bm{\theta},\bm{\phi})$, has a multivariate normal PDF form:
\begin{eqnarray}
L(\bm{\bm{\theta}},\bm{\phi}) &=& \frac{1}{(2\pi)^{(mN)/2}|\Sigma_l|^{1/2}}\exp^{-(\bm{Y-\mu_l})^T\Sigma_l^{-1}(\bm{Y-\mu_l})/2},
\end{eqnarray}
where the mean is:
\begin{eqnarray}
\bm{\mu}_l &=& (\textrm{A}\bm{f}(\bm{\theta},\bm{\phi},k),...,\textrm{A}\bm{f}(\bm{\theta},\bm{\phi},Nk)),
\end{eqnarray}
and the covariance matrix is:
\begin{eqnarray}
\Sigma_l &=& \textrm{U} \otimes \textrm{V} + \sigma^2\textrm{I},
\end{eqnarray}
where $\textrm{U}_{ab} = k \min(a,b)$ with $\textrm{U} \in \mathbb{R}^{N \times N}$, and $\textrm{V} = \textrm{A}\Sigma \textrm{A}^{\intercal}$. Also, the symbol $\otimes$ stands for the Kronecker product. $\bm{Y}_{ck}$ is multivariate normal (conditioning on $\bm{\theta}$ and $\bm{\phi}$) as a direct consequence of equations 7 and 9, noting that $\bm{X}_{ck}$ and $\bm{\eta}_{ck}$ are independent conditional on $\bm{\theta}$ and $\bm{\phi}$. Moreover, the linearity property of expectations can be used to show that the mean of $\bm{Y}_{ck}$ is $E[\textrm{A}\bm{S}_{ck}+\bm{\eta}_{ck}] = E[\textrm{A}\bm{S}_{ck}]+E[\bm{\eta}_{ck}] = E[\textrm{A}\bm{f}(\bm{\theta},\bm{\phi},ck)+\textrm{A}\bm{X}_{ck}]+E[\bm{\eta}_{ck}] = E[\textrm{A}\bm{f}(\bm{\theta},\bm{\phi},ck)]+E[\textrm{A}\bm{X}_{ck}]+E[\bm{\eta}_{ck}] =  \textrm{A}\bm{f}(\bm{\theta},\bm{\phi},ck) + \bm{0} + \bm{0}$ (again, conditional on fixed $\bm{\theta}$ and $\bm{\phi}$ fixed). Appendix A contains more details of the covariance matrix.

Since evaluating Eq. 12 requires the calculation of the inverse of the matrix $\Sigma_l$ and its determinant, these must be calculated efficiently (generally this takes $O(N^3m^3)$ operations, which can grow very quickly with more space and time observations). Since $\textrm{U}^{-1}$ is tridiagonal, the bandwidth of $\textrm{U}^{-1}$ is 1, and the band-limited nature of $\textrm{U}^{-1}$ allows us to compute $\Sigma_l^{-1}$ and $|\Sigma_l|$ in $O(Nm^3)$ time \citep{rue2001fast, golub2012matrix}. More details for this derivation are given in Appendix A. While using band-limited linear algebra routines can improve computation, in the next subsection we derive an approximation to the likelihood that is embarrassingly parallel and can therefore accelerate computation even more. 

\subsection{An approximate likelihood}
Here we show how to approximate the likelihood in a way that leads to embarrassingly parallel computation. The likelihood $p(\bm{Y_k,...,Y_{Nk}}|\bm{\bm{\theta}},\bm{\phi})$ can be equivalently written as $p(\bm{Y_k}|\bm{\bm{\theta}},\bm{\phi})p(\bm{Y_{2k}|Y_k},\bm{\bm{\theta}},\bm{\phi})... \\p(\bm{Y_{Nk}|Y_k,..,Y_{(N-1)k}},\bm{\bm{\theta}},\bm{\phi})$.
First note that:
\begin{eqnarray}
\bm{Y}_{k} &=& \textrm{A}\bm{f}(\bm{\theta},\bm{\phi},k)+\bm{\eta}_{k}+\sum_{j=1}^{k} \textrm{A}\bm{\epsilon}_{j}.
\end{eqnarray}
Hence, $p(\bm{Y}_{k}|\bm{\theta},\bm{\phi})$ is multivariate normal with mean $\textrm{A}\bm{f}(\bm{\theta},\bm{\phi},k)$ and covariance matrix $\textrm{A}(k\Sigma)\textrm{A}^{\intercal}+\sigma^2\textrm{I}$.
More generally, we have the relationship:
\begin{eqnarray}
\label{eq:rec}
\bm{Y}_{ck} &=& \bm{Y}_{(c-1)k} + \textrm{A}[\bm{f}(\bm{\theta},\bm{\phi},ck)-\bm{f}(\bm{\theta},\bm{\phi},(c-1)k)] + \bm{\eta}_{ck} - \bm{\eta}_{(c-1)k} + \sum_{j=(c-1)k+1}^{ck} \textrm{A}\bm{\epsilon}_{j}.
\end{eqnarray}
Thus we can approximate  $p(\bm{Y}_{ck}|\bm{Y}_k,..,\bm{Y}_{(c-1)k},\bm{\theta},\bm{\phi})$ as a MVN distribution with mean $\bm{Y}_{(c-1)k} + \textrm{A}[\bm{f}(\bm{\theta},\bm{\phi},ck)-\bm{f}(\bm{\theta},\bm{\phi},(c-1)k)]$ and covariance matrix $\textrm{A}(k\Sigma)\textrm{A}^{\intercal} +2\sigma^2\textrm{I}$. Nonetheless, to clarify,  $p(\bm{Y}_{ck}|\bm{Y}_k,..,\bm{Y}_{(c-1)k},\bm{\theta},\bm{\phi})$ is not exactly a MVN with mean $\bm{Y}_{(c-1)k} + \textrm{A}[\bm{f}(\bm{\theta},\bm{\phi},ck)-\bm{f}(\bm{\theta},\bm{\phi},(c-1)k)]$ and covariance matrix $\textrm{A}(k\Sigma)\textrm{A}^{\intercal} +2\sigma^2\textrm{I}$ because $\bm{Y}_{(c-1)k}$ and $\bm{\eta}_{(c-1)k}$ are dependent. However, when the magnitude of the observation error $\bm{\eta}_{(c-1)k}$ is much smaller in comparison to the magnitude of the observation $\bm{Y}_{(c-1)k}$, and for $\bm{Z} \sim \textrm{MVN}(0,\sigma^2\textrm{I})$ with $\bm{Z}$ independent of $\bm{Y}_{(c-1)k}$, $\bm{Y}_{(c-1)k} - \bm{\eta}_{(c-1)k}  \approx \bm{Y}_{(c-1)k} - \bm{Z}$. 
 
This approximation is embarrassingly parallel because each of the $N$ terms in the product form of the likelihood $p(\bm{Y}_k,...,\bm{Y}_{T}|\bm{\theta},\bm{\phi}) = p(\bm{Y}_k|\bm{\theta},\bm{\phi})p(\bm{Y}_{2k}|\bm{Y}_k,\bm{\theta},\bm{\phi})...p(\bm{Y}_{Nk}|\bm{Y}_k,..,\bm{Y}_{(N-1)k},\bm{\theta},\bm{\phi})$ (or sum, if computing the log-likelihood) can be evaluated independently of each other. Therefore, in parallel, the computation comes down to evaluating a multivariate normal PDF of dimension $m$ -- this can be done in $O(m^3)$.  

\subsection{Computational complexity summary}
If no attention is paid to the structure of $\Sigma_l$, the cost of evaluating $L(\bm{\theta},\bm{\phi})$ is limited by the evaluation of $\Sigma_l^{-1}$ and $|\Sigma_l|$, which generally takes $O(N^3m^3)$ operations. However, the exact likelihood evaluation can be reduced to $O(Nm^3)$ using band-limited numerical linear algebra. The computational complexity of the approximation is also $O(Nm^3)$ (if no parallelism is used). While an exact likelihood is preferred to an approximation, a benefit of the approximation is that it is embarrassingly parallel -- if parallelized, the time complexity is that of evaluating a multivariate normal PDF of dimension $m$, which is $O(m^3)$. Nonetheless, there also exist parallel versions of sparse Cholesky decomposition, for instance in \citet{Gupta:1994:SPA:602770.602898}. Empirical comparisons of the exact and approximate likelihood computations are presented in Section 4.

\section{Analysis of the model and associated methodology}
The purpose of this section is to motivate the various modeling choices introduced in this paper using the previously described test system from glaciology, both in terms of computational run time and quality of inferences. In particular, we compare a posterior based on an emulator to a posterior based on a numerical PDE solver, motivate the use of the random walk error-correcting process with residual analysis, examine the impact of prior information encoded into the error-correcting process on the bias of posterior distributions for physical parameters, and compare the run-time and accuracy of the likelihood approximation versus the exact likelihood. The physical parameter of interest in these examples is ice viscosity, $B$, whose actual value is the same as \citet{Bueler}, \citet{EISMINT}, and \citet{gopalan2018bayesian}:  $31.7 \times 10^{-25}$ in units of $s^{-1}Pa^{-3}$.

Consistent with \citet{gopalan2018bayesian} is the choice of settings for the numerical PDE solver: a 21 by 21 grid (so $n = 441$) is used with $\Delta_x = \Delta_y = 10^5$ m and $\Delta_t = .1$ years. Note that, consequently, the number of simulator runs (25) is much smaller than the dimensionality of the output of the solver (441). 

\subsection{Posterior inference of the ice viscosity parameter with an emulator compared to a numerical PDE solver}
In this section, we conduct an empirical study to examine how a first-order spatio-temporal emulator (i.e., an emulator based on the method in Appendix B) compares to a numerical solver of the PDE, both in terms of run-time of computations and posterior inference of ice viscosity. While the precise technical details for constructing a first-order spatio-temporal emulator are given in Appendix B, the idea is to approximate the numerical solver output for each time point that there is collected data. To do this, we train an emulator using the following values for ice viscosity: $\{10, 12.5, 15.0,...,70.0\}$ in units of $10^{-25} s^{-1}Pa^{-3}$, a grid of values that is intentionally coarser than the values used for posterior computation, since in this case the emulator must be used for parameter values not in the training set. We used the \texttt{rbenchmark} \citep{rbenchmark} package to benchmark the run-time of the log-likelihood of the model evaluated at the actual parameter value computed with a numerical solver versus a first-order spatio-temporal emulator, using a MacBook Pro early 2015 model with a 2.7 GHz Intel Core i5 processor and 8 GB 1867 MHz DDR3 memory. The emulator version performs 14.5 times faster (.354 seconds for the emulator based log-likelihood versus 5.148 seconds for the numerical solver based log-likelihood). We also generated samples from the posterior distribution of ice viscosity with grid sampling (grid [10,70] inclusive with grid width .50 in units of  $10^{-25} s^{-1}Pa^{-3}$), using both the numerical PDE version and the emulated version. The summary statistics of $10^6$ posterior samples for ice viscosity using both the emulator and numerical solver are given in Table 1. Qualitatively, the summary statistics are similar. 

The principle behind choosing the ice viscosity parameter values in the training set is to fill the space of the support for ice viscosity, but not to choose a grid as fine as the one used for posterior sampling. (Such an approach would be circular, in that the emulator would just be generating predictions inside of the training set.) However, such a heuristic will not be feasible as the number of parameters grows beyond one parameter (the number of design points would need to grow exponentially in the number of dimensions). In such cases, we suggest using other space-filling designs: notably, a latin hypercube design has been used extensively in the computer experiments literature, for instance in \cite{higdon2008computer}.

\begin {table}[H]
\begin{center}
\begin{tabular}{ |c|c|c|c|c|c|c| }
\hline
Test Case & Min & 1st Quartile & Median & Mean & 3rd Quartile & Max \\ \hline
Emulator SIA solver & $15.0$ & $26.5$  & $27.0$ & $27.4$ & $29.0$ & $38.5$\\ \hline
Numerical SIA solver & $15.0$ & $24.5$  & $26.5$ & $26.3$ & $28.0$ & $37.5$\\ \hline
 \hline
  \end{tabular}
\caption{Summary statistics of $10^6$ posterior samples of the ice viscosity parameter using an emulator for the SIA and a numerical solver for the SIA; qualitatively, these posterior samples are similar. Units are in $10^{-25} s^{-1}Pa^{-3}$.}
\end{center}
\end {table}

\subsection{Assessing a random walk for representing model discrepancy}
The choice of using a random walk to correct for deviations between the output of a computer simulator and the actual physical process values has a few important motivations:

\begin{enumerate}
    \item The inaccuracy of a spatio-temporal computer simulation is most likely going to increase as it is run further into the future. Conveniently, a random walk's variance increases with time -- for example a RW(1) has marginal variance $j\Sigma$ at time $j$.
    \item As shown in Appendix A and Section 3.1, the likelihood involves band-limited matrices, for which there exist specialized numerical linear algebra routines. However, there is a trade-off in bandwidth and the order of the random walk utilized.
    \item Spatial correlations in the inaccuracies of a computer simulation can be captured with the covariance matrix $\Sigma$.
\end{enumerate}

In addition to these motivations, the purpose of this section is to empirically assess how a random walk model performs for correcting the output of a numerical SIA PDE solver. To do this, we use the analytical SIA solution as a gold standard. This is a simplification in the sense that the real glacial dynamics will not follow the SIA PDE and therefore the analytical SIA solution exactly, but nonetheless this is a way to check the veracity of the random walk error model in some capacity -- at the very least, as a model for numerical error but not model uncertainty.

Figure 3 displays the differences between the analytical SIA PDE solution for glacial thickness and the numerical SIA PDE solution for glacial thickness at all of the glacier grid points, run forward for 5000 time steps (i.e., 500 years). More precisely, the points in blue are at the margin of the glacier, the points in red are at the interior, and the points filled in black are close to the top (also referred to as the dome) of the glacier. Recall from Figure 1 that the glacier looks like a shallow ellipsoid sliced in half (in the x-y plane), and so the panel on the right of this figure is a top view of the glacier grid points, which looks like a circle of radius 750 km projected onto the x-y plane. In comparison, the height is 3600 m.

\begin{figure*}[h!]
  \centering
    \includegraphics[width=0.5\textwidth]{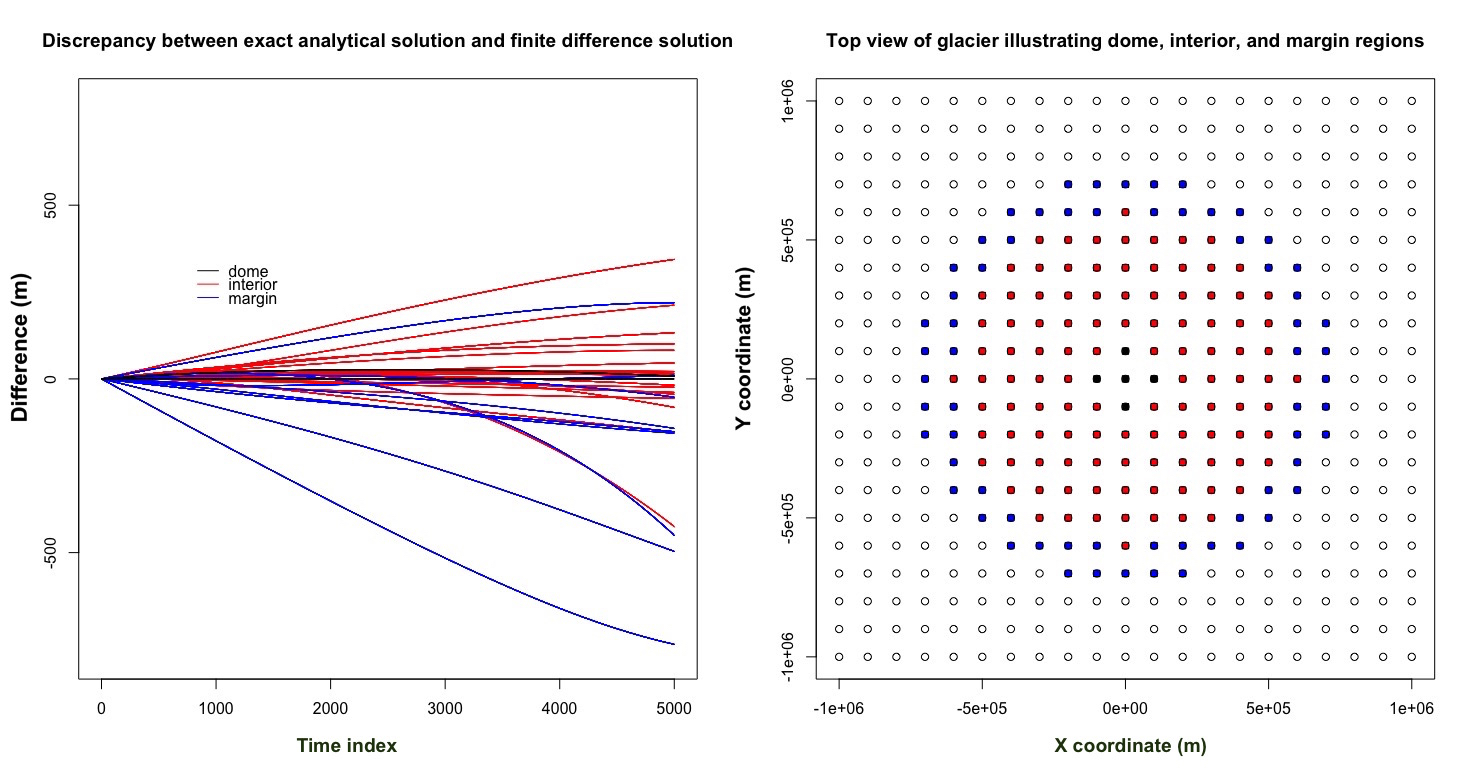}
\caption{An illustration of the difference between the exact analytical solution and the numerical solution for the SIA PDE. On the \textit{right} panel is a top view of the glacier, whose shape looks like a dome, and therefore the projection on to the x-y plane is a circle. The {\color{blue}blue points} signify the margin of the glacier (where it drops down to zero thickness), the {\color{red}red points} are at the interior of the glacier, and the {\color{black}black points} are towards the top of the glacier. The points that are not filled in signify the border of the glacier, where there is no ice thickness. On the \textit{left} panel the discrepancies between the analytical SIA PDE solution and the numerical SIA PDE solution for all grid points are shown. Specifically, the color of each path corresponds to the grid points on the right panel. Additionally, the paths are shown for 500 years, or 5000 time steps.
}
\end{figure*}

The differences are all very smooth (i.e., continuous) functions of time, implying that the numerical SIA PDE solver is producing continuous output as well -- we know that the analytical solution is continuous based on the functional form in Eqs. 3-5. Thus, it appears that a random walk of at least a few orders is necessary to represent these differences. Moreover, as expected from \citet{Bueler}, the largest errors occur at the margin, whereas the interior and dome differences are less extreme.

To assess if a random walk model is appropriate, for each time point $j$ and for orders 1-7, we computed residuals, in other words, the left hand side of Eq. 8, which should theoretically be distributed like $\bm{\epsilon_j}$ (i.e., independent $\textrm{MVN}(0, \Sigma)$ random variables). To compute $\bm{X}_j$, we take the difference $\bm{S}_j-\bm{f}(\bm{\theta},j)$, where $\bm{S}_j$ is the analytical glacial thickness solution to the SIA PDE at time $j$ (i.e., the real physical process for the purpose of this analysis), and $\bm{f}(\bm{\theta},j)$ is the numerical glacial thickness solution to the SIA PDE at time $j$. We examine the residuals for two randomly selected grid points of the glacier (one at the interior and one at the margin) in Figures 4 and 5.

\begin{figure*}[h!]
  \centering
    \includegraphics[width=.6\textwidth]{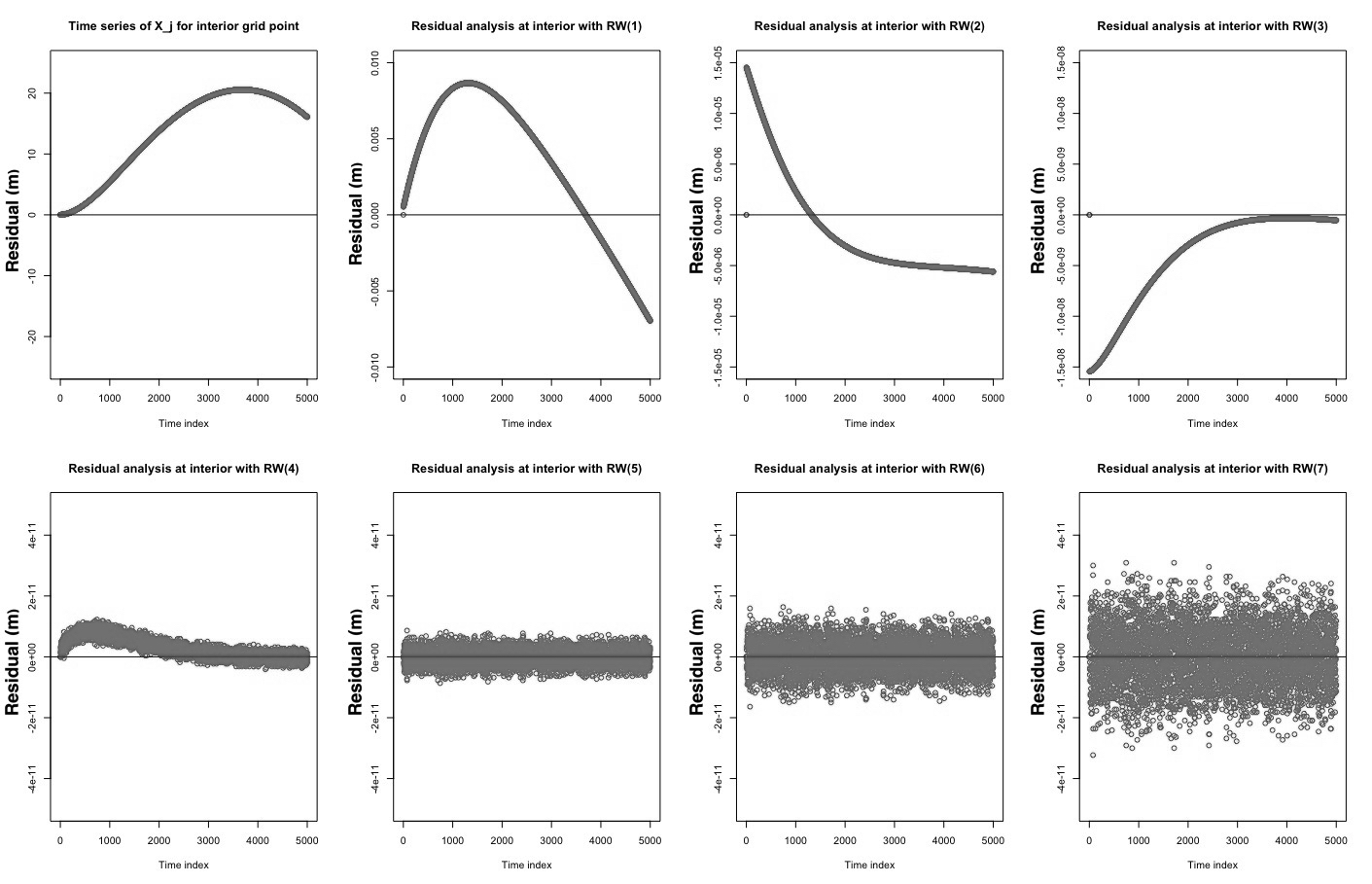}
\caption{This figure displays residuals in units of meters (i.e., the term $\bm{\epsilon}_j$ in Eq. 8) for RW(q) of orders 1-7 for a randomly selected interior grid point. The first four panels display values on different scaled y-axes to better show the shapes, whereas the bottom four panels have the same scaling for the y-axis to be able to compare across the figures. RW(5) and above look like white noise processes, though RW(5) has the smallest variance. 
}
\end{figure*}

\begin{figure*}[h!]
  \centering
    \includegraphics[width=.6\textwidth]{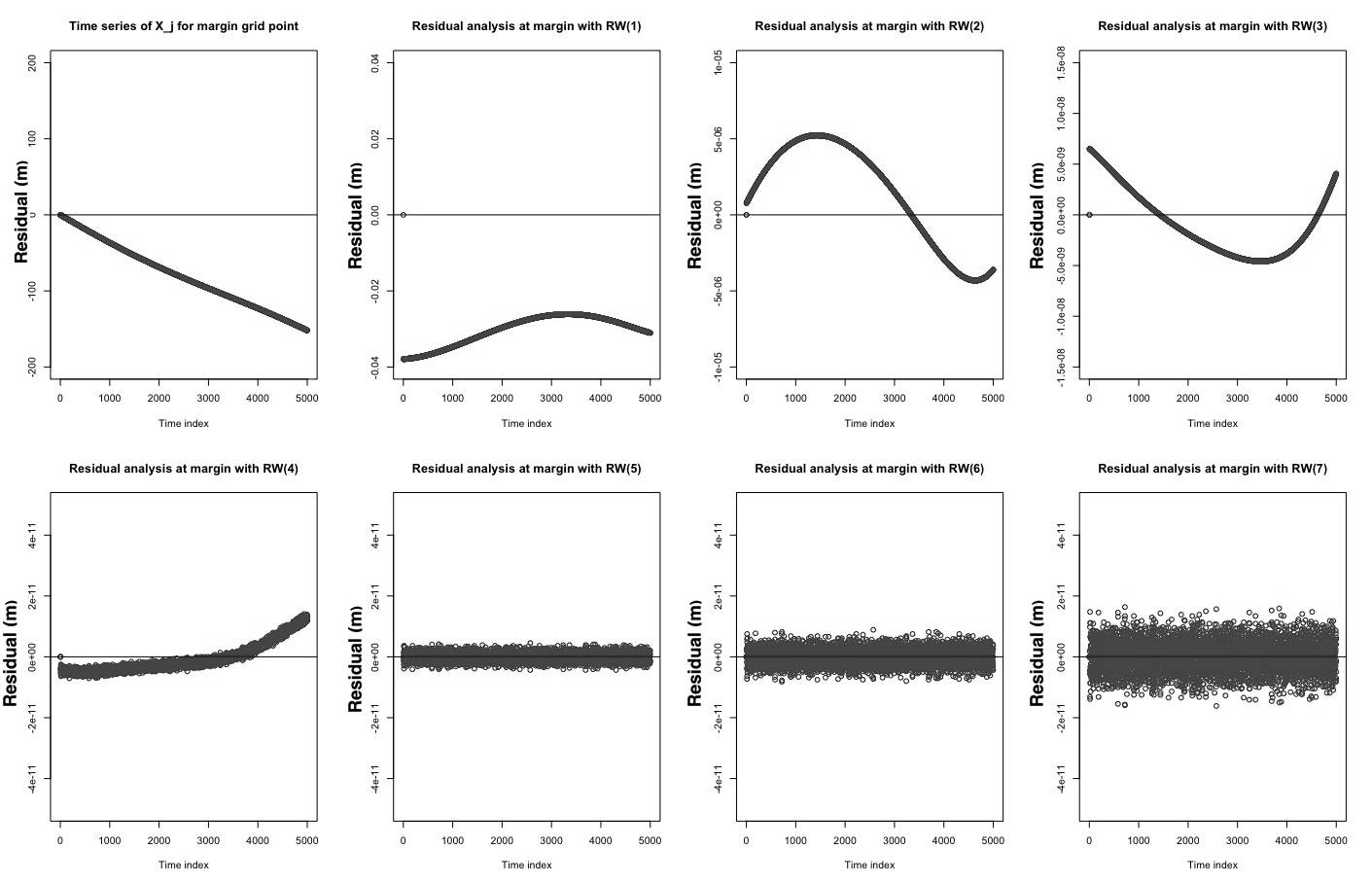}
\caption{This figure also displays residuals in units of meters (i.e., the term $\bm{\epsilon}_j$ in Eq. 8) for RW(q) of orders 1-7 for a randomly selected margin grid point. Just like the previous figure, the first four panels display values on different scaled y-axes to better show the shapes, whereas the bottom four panels have the same scaling for the y-axis to be able to compare across the figures. Just as in the previous figure, RW(5) and above look like white noise processes, though RW(5) has the smallest variance.
}
\end{figure*}

A few important observations should be emphasized based on the empirical analysis displayed in these figures. The first is that a single order random walk substantially filters the discrepancy; for the interior grid point, it is reduced from the order of 10 m to the order of .01 m (1000 times reduction in magnitude), and for the margin grid point from the order of 100 m to .05 m (more than 1000 times reduction). Additionally, for both the interior and margin grid points, it appears that RW(5) is optimal in the sense that the residuals  closely resemble a white noise process and have the smallest variance. While the residuals from higher order RW processes also resemble white noise, the magnitude of the noise is larger. Nonetheless, we believe that real physical processes will not always be as smooth as the analytical SIA PDE solutions, and hence it is likely that a lower order RW process will be preferred for real scenarios. 

\subsection{Reducing bias for the posterior distribution of $\bm{\theta}$}
In \citet{discrepancy}, when prior information about the model discrepancy term is introduced in a simple physical system (i.e., a constrained GP over a space of functions), the bias of a posterior distribution of a physically relevant parameter reduces. We have found that a very similar phenomenon occurs in the glaciology test case, a result that was pointed out in \citet{gopalan2018bayesian}. Specifically, in \citet{Bueler}, it is shown that there is large spatial variation in the scale of deviations between the exact solution to the SIA and a numerical finite difference solver of the SIA. Specifically, there is spatial variation between the dome, interior, and margin of a glacier, with deviations at the margin being markedly larger than at the interior and dome. To investigate the effect of such prior information, we choose the matrix $\Sigma$ to be such that it is block diagonal with 3 blocks, $\Sigma_{int}$, $\Sigma_{dome}$, and $\Sigma_{margin} $. Each of these blocks is derived from a square exponential covariance kernel with the same length-scale parameter $\bm{\phi}$ = 70 km, but differing variance parameters $\sigma^2_{int}$, $\sigma^2_{dome}$, and  $\sigma^2_{margin}$. If we ignore prior information from \citet{Bueler}, we assume that there is an equal prior probability that each of $\sigma^2_{int}$, $\sigma^2_{dome}$, and $\sigma^2_{margin}$ is in the set $\{.1, 1, 10, 100\}$ in units of $m^2$. If we use prior information from \citet{Bueler}, we instead assume equal prior probability on $\{.1,1\}$ for $\sigma^2_{int}$, $\{1,10\}$ for $\sigma^2_{dome}$, and $\{10,100\}$ for $\sigma^2_{margin}$ (again all units are $m^2)$. As shown in \citet{gopalan2018bayesian}, the posterior for ice viscosity is less biased in the case that incorporates prior information for the scale of errors; this phenomenon is explored again in the next section.

While in the above discussion we have not been precise about the term bias, the following ought to make this notion more rigorous. Let $\bm{\theta}_0$ be the true parameter, and $\hat{\bm{\theta}}$ be an estimator of $\bm{\theta}_0$. The frequentist definition of bias is usually $E[\hat{\bm{\theta}} - \bm{\theta}_0]$, where the expectation (i.e., average) is taken over the sampling distribution, $p(\bm{Y}|\bm{\bm{\theta}}_0)$. The Bayesian notion of bias used informally in the preceding paragraph (and essentially the same notion as in \citet{discrepancy}) is $b(\bm{Y},\bm{\bm{\theta}}_0) = E[\bm{\bm{\theta}} - \bm{\bm{\theta}}_0]$, where the expectation (i.e., average) is taken with respect to the posterior distribution of $\bm{\bm{\theta}}$, $p(\bm{\theta}|\bm{Y})$. Consider $E[b(\bm{Y},\bm{\bm{\theta}}_0)]$, where the (outer) expectation is taken with respect to the sampling distribution. Then $E[b(\bm{Y},\bm{\bm{\theta}}_0)] = E[E[\bm{\bm{\theta}}-\bm{\bm{\theta}}_0]] = E[E[\bm{\bm{\theta}}]-\bm{\bm{\theta}}_0] = E[\hat{\bm{\bm{\theta}}}-\bm{\bm{\theta}}_0]$, which is the frequentist bias. In other words, the frequentist bias is equivalent to the average of $b(\bm{Y},\bm{\bm{\theta}}_0)$ over the sampling distribution, if the posterior mean is chosen as an estimator. In the glaciology test case, we have (informally) not noticed much variability in the posterior for ice viscosity over repeated sampling of the data, and hence the distinction between Bayesian bias and frequentist bias is not significant.

The reader may wonder why a fixed $\bm{\theta}_0$ was assumed in the preceding paragraph, despite that a Bayesian model has been presented in this paper. In fact, it is typical to assume that the actual value of a parameter is fixed, despite ascribing a probability distribution to it in the form of a prior or posterior. Conceptually, such a probability distribution is a representation of a modeler's uncertainty regarding the fixed, unknown value of the parameter. For more on this interpretation of Bayesian statistics, the reader can consult results of statistical decision theory (e.g., on admissibility) in \cite{lehmann2003theory} and \cite{robert2007bayesian}. This viewpoint is also taken in Bayesian asymptotic analysis, such as the Bernstein-von Mises theorem \citep{van2000asymptotic, shen2001}.

\subsection{Inferring $\Sigma$}

The covariance matrix $\Sigma$, first introduced after Equation 7 in Section 3, determines the spatial correlation inherent in the error-correcting process, $\bm{X}$. Since spatial correlation in the error-correcting process is important to model (which is particularly evident in the glaciology example of \citet{Bueler}), we need to discuss how $\Sigma$ ought to be specified. Choosing $\Sigma$ can be difficult if no or little prior information is available, and in such a case, we suggest:

\begin{eqnarray*}
\Sigma &=& \textrm{diag}(\bm{v})\textrm{R} \ \textrm{diag}(\bm{v}), 
\end{eqnarray*}
where $\textrm{log}(\bm{v}) \sim \textrm{MVN}(\bm{\mu}_v,\Sigma_v)$,  $\Sigma_v$ is derived from a GP kernel such as squared-exponential or Mat\'{e}rn kernel, and $\textrm{R}$ is a correlation matrix also derived from a GP kernel. To avoid non-identifiability and complexity of inference, it is suggested to pre-specify the parameters of these GP kernels.  This approach is similar to the modeling strategy employed in \citet{doi:10.1002/env.2343}. The intuition behind this approach is that the term $\bm{v}$ encodes spatial variability in the scale of deviations between the output of a computer simulator and the true physical process, and spatial correlation in these deviations is strongly enforced with non-diagonal terms in both $\Sigma_v$ and $\textrm{R}$.

Figure 6 illustrates a map of the mean posterior field for the variances of the error-correcting process, where the area of each circle is proportional to the inferred posterior mean of variance; due to a multivariate normal prior on $\textrm{log}(\bm{v})$, elliptical slice sampling is used as the method for posterior sampling \citep{murray2010elliptical}. Consistent with \citet{Bueler}, the variances tend to increase at the margins and are smaller at the interior. Additionally, the scaled differences between the analytical solution and numerical solver at the final time point the simulator is run (where scaling is inverse of the posterior mean of standard deviation) should theoretically approach a mean zero normal distribution according to the model. The p-value for an Anderson-Darling test is .436, suggesting that the scaled differences between the analytical solution and numerical solver are consistent with a normal distribution. Moreover, the sample mean for these scaled differences is .079 and the sample standard deviation is .409.

\begin{figure*}[h!]
  \centering
    \includegraphics[width=0.5\textwidth]{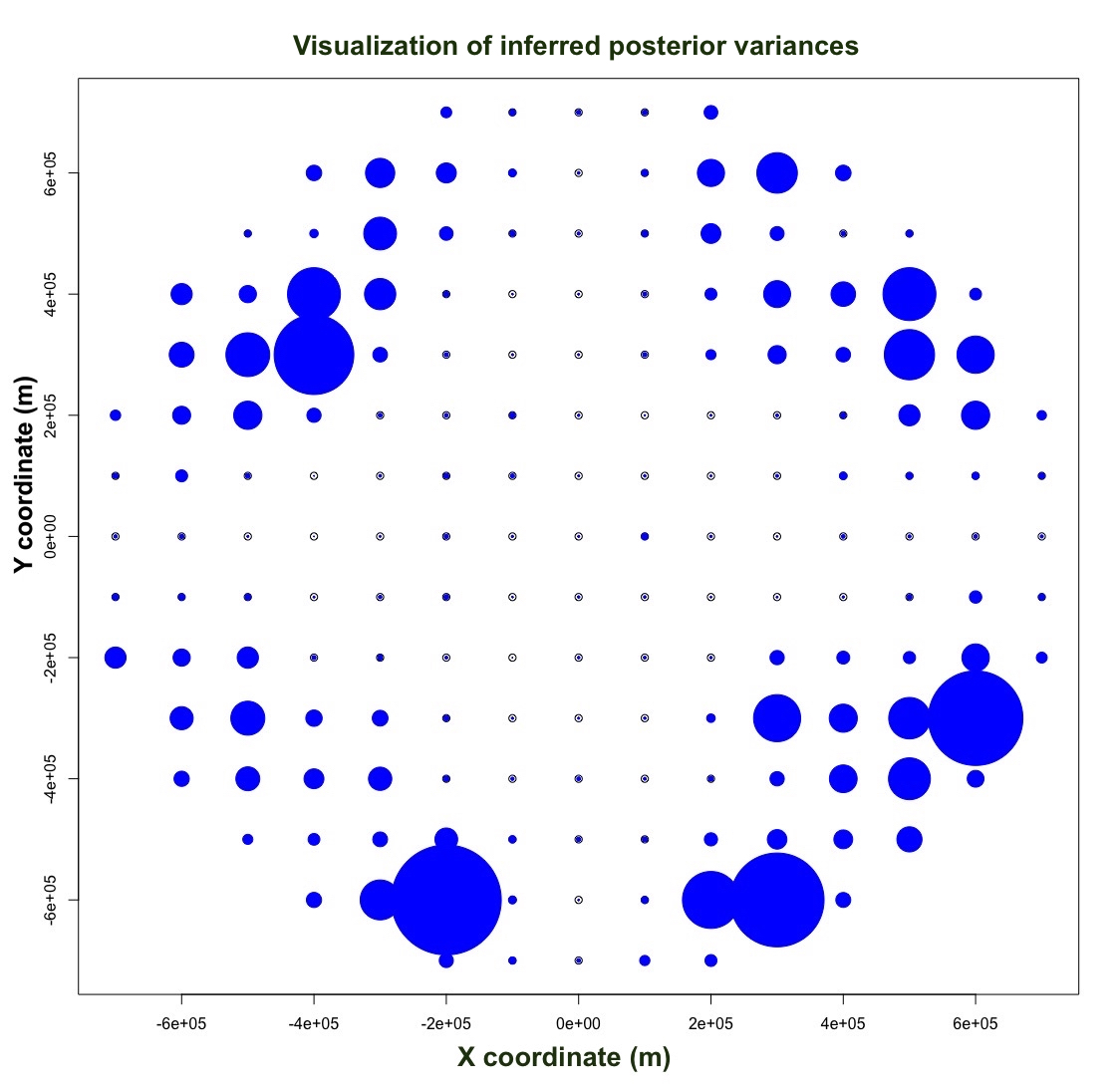}
\caption{Inferred posterior variance field of the error-correcting process, where the area of each circle is proportional to the variance at the grid point centered at the circle. Qualitatively, this field behaves as one would expect from the the work of \citet{Bueler}, where the authors demonstrate that numerical inaccuracies for the SIA PDE are greatest toward the margin, but much smaller at the interior of the glacier.}
\end{figure*}

As is discussed in the previous subsection, prior information for $\Sigma$ has an effect on the inference of physical parameters (i.e., ice viscosity), and in particular, a lack of prior information can lead to a very biased posterior distribution for physical parameters.  To compare the fitted $\Sigma$ using a GP field against the $\Sigma$ matrices discussed in the previous section, we show in Table 2 a comparison of posterior inference for the ice viscosity parameter for three choices of $\Sigma$. The first choice of $\Sigma$ is the posterior mean of samples assuming the structure $\Sigma = \textrm{diag}(\bm{v})\textrm{R} \textrm{diag}(\bm{v})$, with $\textrm{log}(\bm{v}) \sim \textrm{MVN}(\bm{\mu_v},\Sigma_v)$. In the second and third scenarios, $\Sigma$ is block diagonal with three variance parameters for each of the three blocks. A weakly informative case assumes that $\sigma^2_{int} = \sigma^2_{dome} = \sigma^2_{margin} = .1$, whereas a more informative case (using prior information from \citet{Bueler}) has $\sigma^2_{int} = \sigma^2_{dome} = .1$ and $\sigma^2_{margin} =  10$ (all units are $m^2$). The scenario for weak prior information for $\Sigma$ results in a very biased posterior distribution whose support does not cover the actual parameter value ($31.7 \times 10^{-25}$ in units of $s^{-1}Pa^{-3}$) -- the maximum in this case is $26.5 \times 10^{-25}$ in units of $s^{-1}Pa^{-3}$.  While the (absolute) biases of the posterior for ice viscosity for the GP field version compared to the prior information from \citet{Bueler} are comparable (5.09 versus 4.01 in units of  $10^{-25} s^{-1}Pa^{-3}$), the posterior variance is markedly larger in the former case. This result suggests that prior knowledge from a domain expert is likely to be useful in determining $\Sigma$, though in a case when that does not exist, the methodology described in this section is an adequate alternative.

\begin {table}[H]
\begin{center}
\begin{tabular}{ |c|c|c|c|c|c|c| }
\hline
Test Case & Min & 1st Quartile & Median & Mean & 3rd Quartile & Max \\ \hline
$\Sigma$ with GP field& $10.0$ & $21.0$  & $36.0$ & $35.7$ & $50.5$ & $70.0$\\ \hline
$\Sigma$ with strong prior information & $18.0$ & $25.0$  & $26.5$ & $26.6$ & $28.0$ & $35.5$\\ \hline
$\Sigma$ with weak prior information & $12.5$ & $18.5$  & $19.5$ & $19.5$ & $20.5$ & $26.5$\\ \hline
 \hline
  \end{tabular}
\caption{Summary statistics of $10^6$ posterior samples of the ice viscosity parameter under three versions of $\Sigma$. While the weakly-informative case leads to a very biased posterior, the biases for the ice viscosity posterior in the first two $\Sigma$ matrices are comparable. Nonetheless, the posterior variance is much less in the case with prior information from \citet{Bueler}.}
\end{center}
\end {table}

\subsection{Exact versus approximate likelihood}
In Section 3.1, we showed an exact way to calculate the model likelihood as well as an approximation in Section 3.2. In this subsection, our purpose is to compare these two methods of likelihood computation in terms of run-time and posterior inference. Using a MacBook Pro early 2015 model with a 2.7 GHz Intel Core i5 processor and 8 GB 1867 MHz DDR3 memory (as before), one component of the log-likelihood approximation (which can be computed in an embarrassingly parallel fashion with the other components of the sum) takes .0179 s, whereas the full log-likelihood calculation, as in Section 3.1, is .354 seconds (in both cases, using a first-order emulator). The results of comparing posterior samples for the ice viscosity parameter are given in Table 3 -- thus, while the mean, median, first, and third quartiles are comparable, the approximate version has larger posterior uncertainty than the exact version as is evidenced by the wider tails. These results suggest that, while there is likely a computational speed-up afforded by using the approximation (i.e., at least an order of magnitude), the price to pay is increased posterior uncertainty.
\begin {table}[H]
\begin{center}
\begin{tabular}{ |c|c|c|c|c|c|c| }
\hline
Test Case & Min & 1st Quartile & Median & Mean & 3rd Quartile & Max \\ \hline
Exact likelihood & $15.0$ & $26.5$  & $27.0$ & $27.4$ & $29.0$ & $39.5$\\ \hline
Likelihood approximation & $10.0$ & $26.0$  & $28.0$ & $27.7$ & $31.0$ & $52.5$\\ \hline
 \hline
  \end{tabular}
\caption{Summary statistics of $10^6$ posterior samples of the ice viscosity parameter using an exact likelihood and a likelihood approximation (units are in $10^{-25} s^{-1}Pa^{-3}$). While the 1st quartile, median, mean, and 3rd quartile are similar, the tails in the approximation are much wider.}
\end{center}
\end {table}

\section{Generality of the model and methodology}
Though we have tested the model and methodology in the previous sections in the context of a glaciology example, it should be noted that they can be used in other physical systems with similar components. In essence, this modeling and methodology can be applied in scenarios where:
\begin{enumerate}
\item A computer program (i.e., \textit{computer simulator}) is available to simulate a continuous physical process through space and time, but there is a deviation between the output of the computer simulator and the actual physical process.
\item The deviations between the computer simulator output and the actual physical process values tend to grow with time and exhibit spatial correlation structure.
\item Measurements of the physical process are available, but they are potentially scarce both in space and time.
\item Physical parameters governing the physical process are uncertain but can be constrained with domain knowledge for the random walk error covariance (i.e., $\Sigma$).
\end{enumerate}
Recall that at the process level, the model stipulates that:
\begin{eqnarray}
\bm{S}_{j} &=& \bm{f}(\bm{\theta},\bm{\phi},j) + \bm{X}_j.
\end{eqnarray}
To apply the same setup to another physical scenario, a different version of $\bm{f}(.,.,.)$, such as a numerical PDE solver for another system of spatio-temporal PDEs besides the SIA, can be used. However, while $\bm{f}(.,.,.)$ will need to be tailored to another physical scenario based on a different numerical scheme or physical model, the $\bm{X}_j$ term would be modeled in the same way (i.e., with a random walk).

\section{Conclusion}
The objective of this work has been to set forth a versatile physical-statistical model in the Bayesian hierarchical framework that incorporates a computer simulator for a physical process, such as a numerical solver for a system of PDEs. Posterior inference for physical parameters (and, consequently, posterior predictions of the physical process) can be computationally demanding within this model, since each evaluation of the likelihood requires a full PDE solve and computing the inverse and determinant of a large covariance matrix. Therefore, we have set forth two main ways to speed up computation: first is the use of bandwidth limited linear algebra in a manner similar to \citet{rue2001fast} for quickly handling the covariance matrix in the likelihood, and the second is the use of spatio-temporal emulation in a manner similar to \citet{Hooten2011} to emulate a PDE solver that is expensive to evaluate. An additional method for speeding up computation is to approximate the likelihood in a way that leads to embarrassingly parallel computation. The utility of this model and corresponding inference methodology is demonstrated with a test example from glaciology.

A unique feature of this work is how we represent the discrepancy between a computer simulator for a physical process and the real physical process values. One approach, as in  \citet{kennedy2001bayesian} and \citet{discrepancy}, is to assume that this is a fixed yet unknown function that can be learned with a GP (or constrained GP) prior distribution over a space of functions. Instead, we assume that this discrepancy is a spatio-temporal stochastic process (i.e., a random walk), which is motivated by the fact that a computer simulation is likely to become less accurate as it is run forward in time, as well as exhibit some degree of spatial correlation in inaccuracies. An interesting consequence of this modeling decision is that linear algebraic routines for band-limited matrices can be utilized for evaluating the likelihood of the model in an efficient manner. Another interesting artifact of this approach is that when prior information is used for the random walk's error term (i.e., in $\Sigma$), the bias for the posterior distribution of $\bm{\theta}$ is reduced. The same phenomenon is exhibited in the work of \citet{discrepancy}, where a constrained GP prior over a space of functions ends up reducing the bias of the physical parameter posterior distribution. 

Despite that the model and methodology appear to perform well in the analysis of this paper, it is important to comment on some potential drawbacks of the approach, particularly when applied to other physical contexts. In this paper, emulation works adequately with a single parameter, though emulators do not always work well in other applications or higher dimensional parameter spaces. For example, \cite{doi:10.1080/01621459.2018.1514306} document some shortcomings of a principal components based emulator in climate modeling. The second main computational advantages stem from log-likelihood evaluation speed-ups. The use of bandwidth limited matrix algebra for the exact log-likelihood can be used so long as the model holds, which may not always be the case (e.g., with a non-Gaussian data distribution). Additionally, the log-likelihood approximation holds when the measurement errors are small relative to the signal modeled, which depends on the measurement instruments used to collect the data. For instance, on common geophysical scales of thousands of meters, light detection and ranging (LIDAR) or digital-GPS data have maximum errors on the order of a meter.

Additionally, if it is not possible to program the computer simulator to produce output at the data measurement locations, there are essentially two main ways to handle such a scenario. The first is to use spatial kriging to predict the value of the computer simulator at the spatial locations where data are collected, given the output of the computer simulator at the grid points. A simpler approach is to use inverse-distance weighting of the simulator output at the nearest neighbors; that is, take a weighted average of the four nearest grid points of the simulator, where the weights are proportional to the inverse of distance. Such an approach, for example, has been used in \citet{doi:10.1002/env.2343}.

Future research will include predicting Langj\"{o}kull glacier surface elevation using the modeling and methodology within this paper, based on actual data collected by the UI-IES. 

\appendix
\section{Derivation of the exact likelihood and computational simplifications}
As was shown in Appendix B of \cite{gopalan2018bayesian}, the covariance matrix for the observed data can be written as $\textrm{U} \otimes \textrm{V} + \sigma^2\textrm{I}$, where $\textrm{U}_{ab} = k \min(a,b)$ with $\textrm{U} \in \mathbb{R}^{N \times N}$, and $\textrm{V} = \textrm{A}(\Sigma)\textrm{A}^{\intercal}$. It can be verified that $\textrm{U}^{-1}$ is tridiagonal, so it has bandwidth one -- more specifically:

\begin{eqnarray}
\textrm{U}^{-1} &=& k^{-1}
\begin{bmatrix}
    2   & -1 & 0 & \dots &  &  \\
    -1  & 2 & -1 & 0 & \dots &  \\
    0   & -1 & 2 & -1 & 0 & \dots & \\
    0   & 0 & \ddots & \ddots & \ddots\\
    0 & ... & &-1 & 2 & -1 \\
    0 & ... & & &  -1 & 1
\end{bmatrix}.
\end{eqnarray}

One useful property of the Kronecker product is that $(\textrm{U} \otimes \textrm{V})^{-1} = \textrm{U}^{-1} \otimes \textrm{V}^{-1}$. Therefore:
\begin{eqnarray}
(\textrm{U} \otimes \textrm{V})^{-1} &=& \textrm{U}^{-1} \otimes \textrm{V}^{-1} \\
                   &=& k^{-1}
\begin{bmatrix}
    2\textrm{V}^{-1}   & -\textrm{V}^{-1} & 0 & \dots &  &  \\
    -\textrm{V}^{-1}  & 2\textrm{V}^{-1} & -\textrm{V}^{-1} & 0 & \dots &  \\
    0   & -\textrm{V}^{-1} & 2V^{-1} & -\textrm{V}^{-1} & 0 & \dots & \\
    0   & 0 & \ddots & \ddots & \ddots\\
    0 & ... & &-\textrm{V}^{-1} & 2\textrm{V}^{-1} & -\textrm{V}^{-1} \\
    0 & ... & & &  \textrm{V}^{-1} & \textrm{V}^{-1} 
\end{bmatrix},
\end{eqnarray}
whose bandwidth is $O(m)$.

Let us denote $\textrm{U} \otimes \textrm{V}$ as $\textrm{W}$. By the matrix inversion lemma, it follows that $(\sigma^2\textrm{I}+\textrm{W})^{-1} = \sigma^{-2}\textrm{I}-\sigma^{-2}(\textrm{W}^{-1}+\sigma^{-2}\textrm{I})^{-1}\textrm{I}\sigma^{-2}$. The matrix $\textrm{W}^{-1}+\sigma^{-2}\textrm{I}$ has bandwidth $O(m)$ since $\textrm{W}^{-1}$ has bandwidth $O(m)$ as shown previously, so this expression can be computed in $O(Nm^3)$  \citep{rue2001fast,golub2012matrix}.
 
 Similarly, by the matrix determinant lemma, $\textrm{log}[\textrm{det}(\sigma^2\textrm{I}+\textrm{W})]$ is $\textrm{log}[\textrm{det}(\textrm{I}+\sigma^2\textrm{W}^{-1})\textrm{det}(\textrm{W}^{-1})^{-1}]$ = $\textrm{log}[\textrm{det}(\textrm{I}+\sigma^2\textrm{W}^{-1})]$-$\textrm{log}[\textrm{det}(\textrm{W}^{-1})]$. Since both terms are log-determinants of square matrices of dimension $Nm$ and bandwidth $O(m)$, this can be calculated in $O(Nm^3)$ due to the efficient Cholesky factorization of band-limited matrices \citep{rue2001fast, golub2012matrix}.

\section{First-order spatio-temporal emulators}
In the examples of this paper, the function $\bm{f}(.,.,.)$ (i.e., the computer simulator) can take one of two forms: a numerical PDE solver for the SIA, or an emulator constructed from the numerical PDE solver for the SIA. The numerical method for solving the SIA PDE is as given in \citet{gopalan2018bayesian}, and the emulator is constructed based on the finite difference solver in a manner as suggested in \citet{Hooten2011}, termed first-order emulation.

That is, we start with a set of plausible values for ice viscosity: $\{\theta_1,\theta_2,...,\theta_p\}$ and, for each time point there is collected data $ck$, we store a matrix  $\textrm{M}_{ck}$, where the $q\textrm{-th}$ column of matrix $\textrm{M}_{ck}$ is the output of the numerical solver using parameter value $\theta_q$ after running for $ck$ time steps forward.  Thus, each matrix $\textrm{M}_{ck}$ is of dimension $n$ by $p$, and without essential loss of generality we can assume that the number $n$ is much larger than $p$, and each matrix $\textrm{M}_{ck}$ is of rank $p$.

For each matrix, $\textrm{M}_{ck}$, we compute a singular value decomposition (SVD), $\textrm{U}_{ck}\textrm{D}_{ck}\textrm{V}^{\intercal}_{ck}$. The goal is to find a (vector valued) function $v_{ck}(\theta*)$ such that the emulated output at time ${ck}$ for parameter value $\theta*$ is $\textrm{U}_{ck}\textrm{D}_{ck}v_{ck}(\theta*)$. To find the $q\textrm{-th}$ element of $v_{ck}$, we train a random forest \citep{Breiman:2001:RF:570181.570182,randomForest} with $(\theta_1, (V^{\intercal}_{ck})_{q1}), (\theta_2, (V^{\intercal}_{ck})_{q2}),...,(\theta_p, (V^{\intercal}_{ck})_{qp})$ as training data, where $(V^{\intercal}_{ck})_{q1}$ is the first element of the $q\textrm{-th}$ right singular vector, $(V^{\intercal}_{ck})_{q2}$ is the second element of the $q\textrm{-th}$ right singular vector, and so on. Not all of the right singular vectors need be used in emulation, and a heuristic such as an elbow-scree plot or the randomization procedure of \cite{friedman2001elements} can be used to determine the number of right singular vectors to keep. However, if the number of simulator runs ($p$) is much smaller than the dimensionality of the output ($n$), all of the right singular vectors can be utilized with computational savings, as is done in the experiments of this paper. 

We have assumed the initial conditions and boundary conditions are known, since this is the case in the glaciology problems we have studied, where the boundary condition is that glacial thickness is nonnegative, and the initial glacier profile (i.e., a dome) is known. In general, however, $\bm{\phi}$ may be incorporated into the analysis above by considering $\theta$ and $\bm{\phi}$ jointly. Additionally, a variant is to directly emulate the likelihood function. However, since there is flexibility in the choice of $\Sigma$ (which enters into the likelihood), unless one is set on using a particular value of $\Sigma$, it is sensible to emulate the numerical solver as opposed to retraining a likelihood emulator for each potential choice of $\Sigma$.

\bibliographystyle{jasa3}
\bibliography{references.bib}
\end{document}